*To be submitted*

# Beyond FAIR Data: Instrument Traces for Active and Autonomous Scientific Experimentation

Sergei V. Kalinin[1,2,*], Boris N. Slautin[1,*], Yu Liu[1], Charles Cao[2]

[1] Department of Materials Science and Engineering, University of Tennessee; Knoxville, Tennessee, 37996, USA
[2] Department of Electrical Engineering and Computer Science, University of Tennessee; Knoxville, Tennessee, 37996, USA

sergei2@utk.edu, bslauti1@utk.edu

**Abstract**
Artificial intelligence is turning scientific instruments from passive data sources into active systems in which each observation can change what is measured next. We argue that this transition generates a fourth class of scientific record, the experimental trajectory, complementing the sample provenance, the acquired data and instrument metadata, and the analysis workflows. The resulting Instrument Trace objects should ultimately be synchronized with Sample Trace objects describing the evolving specimen provenance and Decision Trace objects describing the human, algorithmic, or agentic logic that selects subsequent actions. We demonstrate retrospective reconstruction of an Instrument Trace from a longitudinal AFM/PFM archive containing approximately 118,000 timestamped events acquired from 2023 to 2026. Working downward through the hierarchy of instrument use, we recover from conventional saved files alone the material-campaign structure of the sample stream, latent probe and calibration states, session-level experimental complexity and control aggressiveness, a recurrent grammar of experimental decisions, and the composite parameter actions from which recorded tuning is assembled. The same analysis exposes what is missing from conventional archives, including unsaved tuning, failed operations, explicit probe and sample-state identities, complete time accounting, exogenous state, and decision rationale. We therefore propose a prospective architecture for active instruments that records project and campaign hierarchy, synchronized sample/instrument/decision traces, controller identity, errors and recovery, environment, action provenance, common action semantics, safety filtering, and replay. Instrument Traces convert instrument archives from repositories of measurements into longitudinal records of how scientific knowledge was acquired and provide a basis for reproducible autonomy, operator training, predictive maintenance, counterfactual analysis, and federated learning across facilities.

## 1. Introduction

Artificial intelligence increasingly acts as a part of the data generation process rather than merely analyzing data *post hoc*. Machine-learning models select measurements, optimize parameters, control experimental hardware, and are beginning to participate in experiment planning. This creates a demand not only for more data, but for data that retain enough context to support machine interpretation and reuse. The FAIR principles - Findable, Accessible, Interoperable, and Reusable - provide the foundational framework for such stewardship [1]. In mature, standardized domains, large interoperable repositories and reproducible computational pipelines have already enabled qualitatively new modes of discovery, exemplified by AlphaFold and RoseTTAFold in structural biology[2, 3] and by the Materials Project, OQMD, AFLOW, JARVIS, NOMAD, and related high-throughput and data-driven discovery frameworks in materials science [4-11]. Microscopy is less forgiving: the scientific meaning of an image or spectrum often depends strongly on instrument configuration, probe state, calibration, feedback conditions, and acquisition history. Reproducibility therefore requires preservation of data plus instrumental metadata and, increasingly, the computational analysis workflow that transforms raw measurements into scientific results. Open microscopy analysis environments and reproducible code workflows, including Pycroscopy, py4DSTEM, and AtomAI, illustrate this move toward executable and inspectable analysis records [12-15], consistent with broader efforts to treat computational workflows themselves as FAIR research objects [16, 17]. In active experiments, however, even this expanded FAIR record is incomplete because the analysis may change the subsequent experiment itself.

A parallel and much older tradition exists in industrial process monitoring. Shewhart's statistical process control established the use of time-resolved measurements to distinguish stable variation from assignable departures from normal operation [18]. Multivariate process monitoring subsequently used latent-variable models, including principal-component analysis, to represent high-dimensional process histories and detect abnormal operating states [19]. Process-history-based fault diagnosis and condition-based maintenance connected sensor traces to fault identification, prognosis, maintenance, and recovery decisions [20, 21], while process mining demonstrated that realized workflows can be reconstructed directly from time-stamped event logs rather than inferred from prescribed procedures [22]. These traditions show that operational histories can reveal latent states, inefficiencies, precursors to failure, and changes in operating regime. Scientific instruments differ from production assets in one critical respect: their trajectory is often changed deliberately to acquire new information as a part of exploration process. For an active instrument, the question is therefore not only whether the process remains healthy, but how the experiment changes state, which action is selected next, and eventually what evidence motivated that action.

This open-ended nature of exploratory research is particularly distinct for microscopy. Automated and autonomous electron and scanning probe microscopy have been formulated as sequential decision systems in which human operators, instrument-native feedback, probabilistic models, and higher-level agents act on different timescales [23-25]. Active-learning implementations in scanning probe microscopy already select where to measure next, which channel to acquire, or how to interrogate a local response based on information accumulated during the campaign [26, 27], while machine-learning-based probe conditioning, image-quality management, and automated STEM demonstrate increasingly direct links between measurements, latent instrument state, and corrective action [28-31]. Related work has emphasized that microscopy workflows span ideation, orchestration, execution, real-time analysis, storage, and theory-experiment interaction rather than a single post-acquisition analysis step [24, 25]. Community roadmaps for autonomous experimentation have likewise identified data and workflow infrastructure, rather than algorithms alone, as a principal bottleneck [32]. These developments move the scientific record from a stored data product toward a record of the experimental workflow that produced it.

The same process-centric view has begun to emerge in self-driving laboratories, from closed-loop reaction optimization and mobile robotic chemistry to autonomous thin-film discovery [33-37]. ESCALATE introduced machine-readable experiment specification and metadata capture [38], and distributed self-driving-laboratory architectures have represented design-make-test-analyze processes and resource states using dynamic knowledge graphs [39]. Individual autonomous campaigns increasingly expose useful trajectories: CAMEO used Bayesian uncertainty to choose sequential synchrotron measurements [40]; A-Lab analyzed successful and failed synthesis trajectories to improve future synthesis policy [41]; SAMPLE compared the search paths followed by multiple autonomous protein-engineering agents [42]; and AlphaFlow learned multistep reaction programs rather than only static parameter values [43]. In microscopy, machine-learning-based STM tip conditioning provides a direct example in which measurement data were used to infer a latent instrument state and trigger corrective action [31]. Retrospective experimental histories can also generate new science: Nega et al. mined 8,470 robotic perovskite crystallization experiments, identified a humidity dependence, and then tested the resulting trace-water hypothesis with 1,296 targeted interventions [44]. These examples show that experimental trajectories are already scientifically useful, but they are typically analyzed within a single designed campaign or represented as orchestration infrastructure rather than as the heterogeneous longitudinal history of a general-purpose instrument.

Here we introduce the Instrument Trace as the longitudinal record of how a scientific instrument is actually used. We organize instrument use hierarchically into projects, campaigns, sessions, and events and place it alongside a Sample Trace and a Decision Trace. We then ask how much of this structure can be recovered retrospectively from existing instrument files. Using a multiyear AFM/PFM archive, we work downward through this hierarchy: the sample stream and its campaign structure, the latent probe and calibration state of the instrument, session-level complexity and control strategy, the event-level grammar of experimental decisions, and finally the individual control changes from which tuning is assembled. Finally, we use the limitations of the retrospective reconstruction to define what future active instruments should record, including complete timing, failed actions, exogenous state, operator and controller identity, common action semantics, physical safety filters, and decision provenance. Although a complete active-experiment record can in principle be connected through persistent identifiers to sample histories, preparation records, LIMS entries, and other forms of sample provenance, we deliberately restrict the retrospective case study here to the instrument-side trace. This is both a useful test and a practical choice: native acquisition files, timestamps, and instrument metadata are commonly retained inside microscopy facilities, whereas sample histories and sample-specific metadata often remain with individual users and are not consistently available to the facility. The present analysis therefore asks how far one can go using the record that is most broadly and routinely available across shared microscopy infrastructure; Section 4 returns to the additional links that should be captured prospectively [45-47].

## 2. Instrument Trace Framework

The framework starts from the hierarchy of instrument use rather than from the file hierarchy. Files are outputs of an experiment; they are not themselves the experiment. A useful trace must preserve the organizational scale on which scientific continuity, sample change, instrument operation, and decision making occur.

## 2.1 From projects to experimental events

We represent scientific instrument use as Project → Campaign → Session → Event (Figure 1). A project is defined by continuity of the scientific objective and may consist of one visit or many months of iterative work. A campaign is one coherent interaction with a defined sample state, potentially over several sessions. Sample identity and sample state are distinct: cleaning, annealing, irradiation, electrical cycling, or other preparation or stimuli-induced changes can create a new state of the same physical specimen. A session is an operational interval in which instrument activity is effectively continuous, representing single instrument-sample pair. In a prospective system the session should be recorded explicitly; in a retrospective archive it must be inferred from timing. An event is the fundamental time-resolved object and can represent an acquisition, movement, parameter change, calibration, tuning operation, intervention, error, recovery action, or decision. Different descriptors are meaningful at different levels: acquisition duty cycle belongs to sessions or campaigns, change in sample preparation is generally a campaign -level event, whereas iterative sample synthesis defines a project.

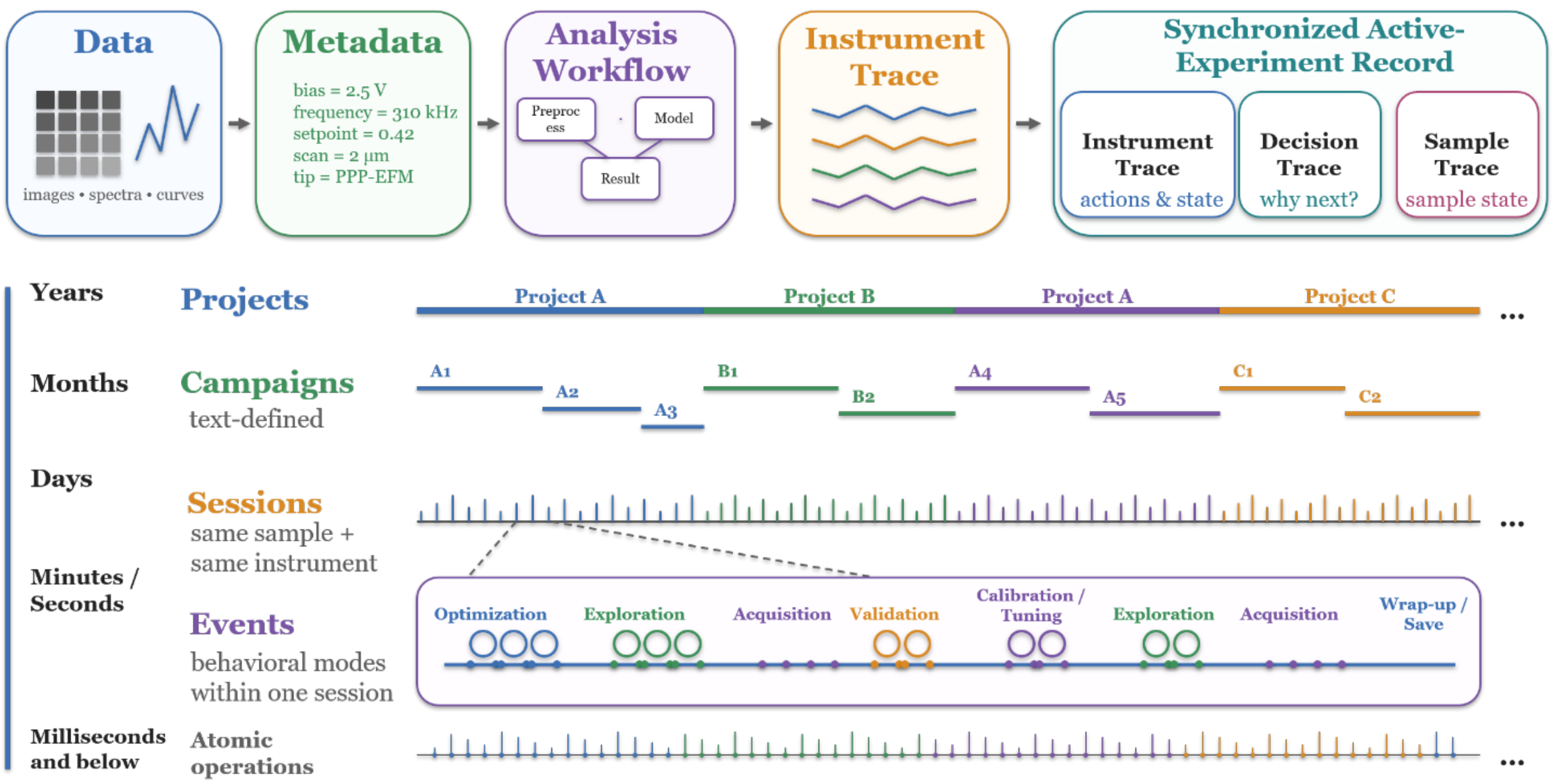

**Figure 1. Instrument Trace framework and hierarchy of experimental activity.** Top: progression from data and metadata through analysis workflows to the Instrument Trace and the synchronized Instrument, Decision, and Sample Trace record. Bottom: nested organization of instrument use from projects and campaigns to sessions, behavioral events, and atomic instrument operations across timescales from years to milliseconds.

## 2.2 Three synchronized traces

The complete record of an active experiment contains three synchronized but differently paced histories. The Sample Trace follows the physical specimen and its modifications. The Instrument Trace follows the measurement system, its probe or interaction state, and the actions performed. The Decision Trace follows the human or algorithmic choices that change what happens next. For scanning probe microscopy, probe state is part of instrument state: nominally identical setpoints or excitation amplitudes have different physical meaning when cantilever stiffness, optical-lever calibration, contact resonance, or tip condition changes. The purpose of synchronization is not to force these processes onto one clock, but to preserve the relationships among a slowly evolving sample, rapid instrument events, and scientific decisions that may operate at project, campaign, or within-campaign scale. Section 3 is organized to follow these three traces in order, and within them to descend the hierarchy of Section 2.1 from the archive as a whole to the individual control change.

## 3. Retrospective Reconstruction of an Instrument Trace

To determine what can be learned without redesigning instrument software or instrument-user record correlation, we deliberately analyze the instrument-side record in isolation. A richer reconstruction could connect instrument events to sample provenance, preparation histories, LIMS records, and user-supplied project metadata through persistent links. However, these records are unevenly available retrospectively: facilities usually retain native acquisition files, timestamps, instrument metadata, and workstation histories, while detailed sample histories often remain with the user. Focusing on the pure Instrument Trace therefore provides both a conservative test of what can be reconstructed from broadly available facility records and a practical baseline for facilities that wish to analyze legacy archives before changing recording practices. For this purpose, we analyzed a longitudinal scanning probe microscopy archive spanning 2023-2026. The index contains 118,298 records, of which 118,296 have valid timestamps, including 32,677 images and 85,619 spectroscopy events. Approximately 1,828 h of non-overlapping acquisition time are recognized. The same general logic can be applied to other active measurement modes, but the state variables must be instrument specific: PFM requires probe, resonance and switching-program descriptors; KPFM would require surface-potential, lift-height or bias-control descriptors; electron microscopy, diffraction and synthesis tools require different physical extensions.

The subsections below proceed from the coarsest recoverable structure to the finest: gross utilization (3A), the sample trace as a sequence of material campaigns (3B), the instrument trace as a sequence of latent probe and calibration states (3C), and the decision trace at three successive scales including session, acquisition, and individual control change (3D–3F). The two state sections precede the three decision sections deliberately, because both establish confounds that must be removed before operator behavior can be interpreted.

### 3A. Gross statistics of instrument use

A 60-min inactivity boundary reconstructs 1,223 sessions, of which 833 contain at least three events and span at least 30 min and are used as substantive sessions for workflow analysis. Recognized acquisition is strongly campaign-like rather than stationary. The most active day contains 23.5 h of acquisition, 50.8% of all recognized acquisition occurs outside 08:00-18:00, and 21.5% occurs on weekends (Figure 2c). These values show extensive unattended use but cannot by themselves identify the controller because the present archive does not persist whether an event was human-, script-, optimizer-, or agent-controlled. The median recognized acquisition fraction within substantive sessions is 47% (interquartile range 32-63%) (Figure 2a,d). Importantly, this is not a full utilization metric: tuning, navigation, rejected acquisitions, reasoning, data transfer, and recovery are incompletely represented in saved files. A future trace therefore needs explicit time accounting rather than treating all non-acquisition time as idle.

To characterize how instrument use changes throughout the week, we represent each active calendar month as a 168-component vector containing acquisition time in each day-of-week/hour bin. A PCA of the unnormalized vectors primarily measures monthly load: its first component explains 21.7 % of the variance and is almost perfectly correlated with total monthly acquisition (Spearman ρ = 0.978). To isolate when the instrument is used rather than simply how much it is used, we normalize each month's 168-component vector by that month's total acquisition time, standardize the 168 features, and perform PCA on the 31 of 36 active months containing at least 5 h of acquisition. PC1 and PC2 explain 12.6% and 9.6% of the normalized schedule variance (Figure 2e). PC1 contrasts Thursday-Friday-Saturday-centered operation against early-week and Sunday-centered operation, whereas PC2 is dominated by weekend/overnight use relative to weekday daytime use (Figure 2f). The month-to-month PCA scores therefore provide a compact representation of changing operating modes rather than a single after-hours fraction.

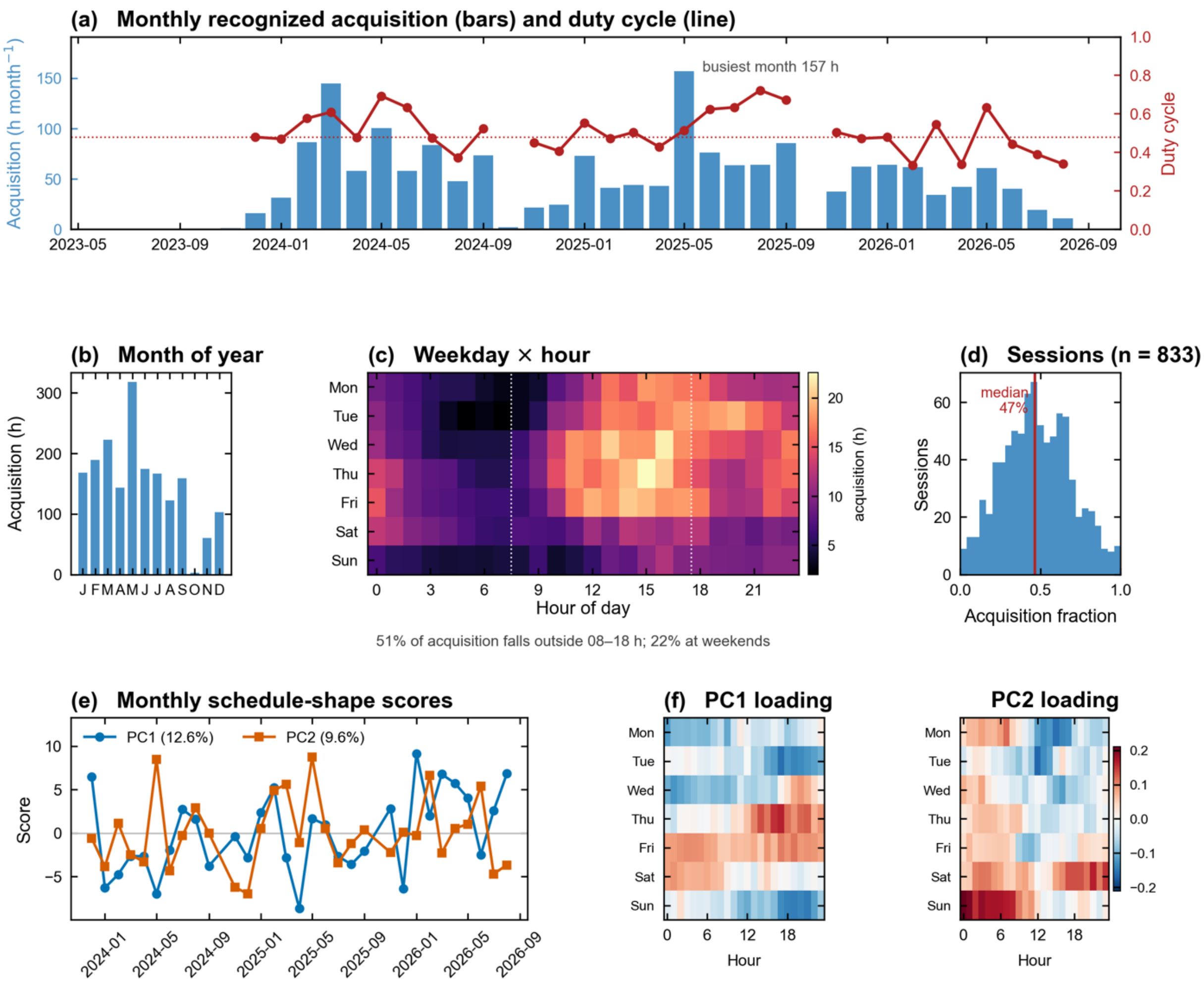

**Figure 2. Gross structure and temporal modes of instrument activity.** (a) Monthly recognized acquisition time (bars) and acquisition duty cycle (line) across the multiyear archive. (b) Total recognized acquisition time pooled by month of year. (c) Distribution of acquisition time across weekday and hour of day; 51% of acquisition occurs outside 08:00–18:00 and 22% on weekends. (d) Distribution of recognized acquisition fraction across 833 substantive sessions; the median is 47%. (e) Monthly scores of the first two principal components of normalized weekly acquisition

profiles, representing changes in the shape of the operating schedule rather than total instrument use. (f) PC1 and PC2 loading maps showing the weekday–hour combinations that define these schedule modes.

### 3B. The sample trace: material campaigns

The coarsest structure above the archive itself is the sample stream, which folder naming resolves into recognizable material campaigns. Lead-based perovskites (PZT and PZTO variants) dominate at 41,316 acquisitions (219 hours) combined, followed by hafnia- and nitride-based systems (140.7 h and 96.8 h, respectively) and samarium-doped bismuth ferrite (40.6 h). The decisive observation is not the ranking but the timing: campaigns arrive in blocks of weeks to months rather than being interleaved (Figure 3d,e). Importantly, only about 20% of the acquired records could be assigned to a specific material system. The remaining ~80% contain no material-identifying information in the metadata and therefore remain unclassified. Material and epoch are therefore confounded across the archive, and an apparent time trend in any acquisition parameter may simply be a change of sample. This affects every pooled statistic in the sections that follow, and it cannot be corrected retrospectively because folder names are operator-chosen and leave roughly four-fifths of acquisitions unresolvable – a direct argument for recording campaign and sample-state identity prospectively or integrating Instrument Trace with sample provenance from laboratory information management systems (LIMS) or other forms of sample provenance recording. Accordingly, the retrospective analysis here remains deliberately restricted to the Instrument Trace, the facility-side record that is most consistently available across shared microscopy environments.

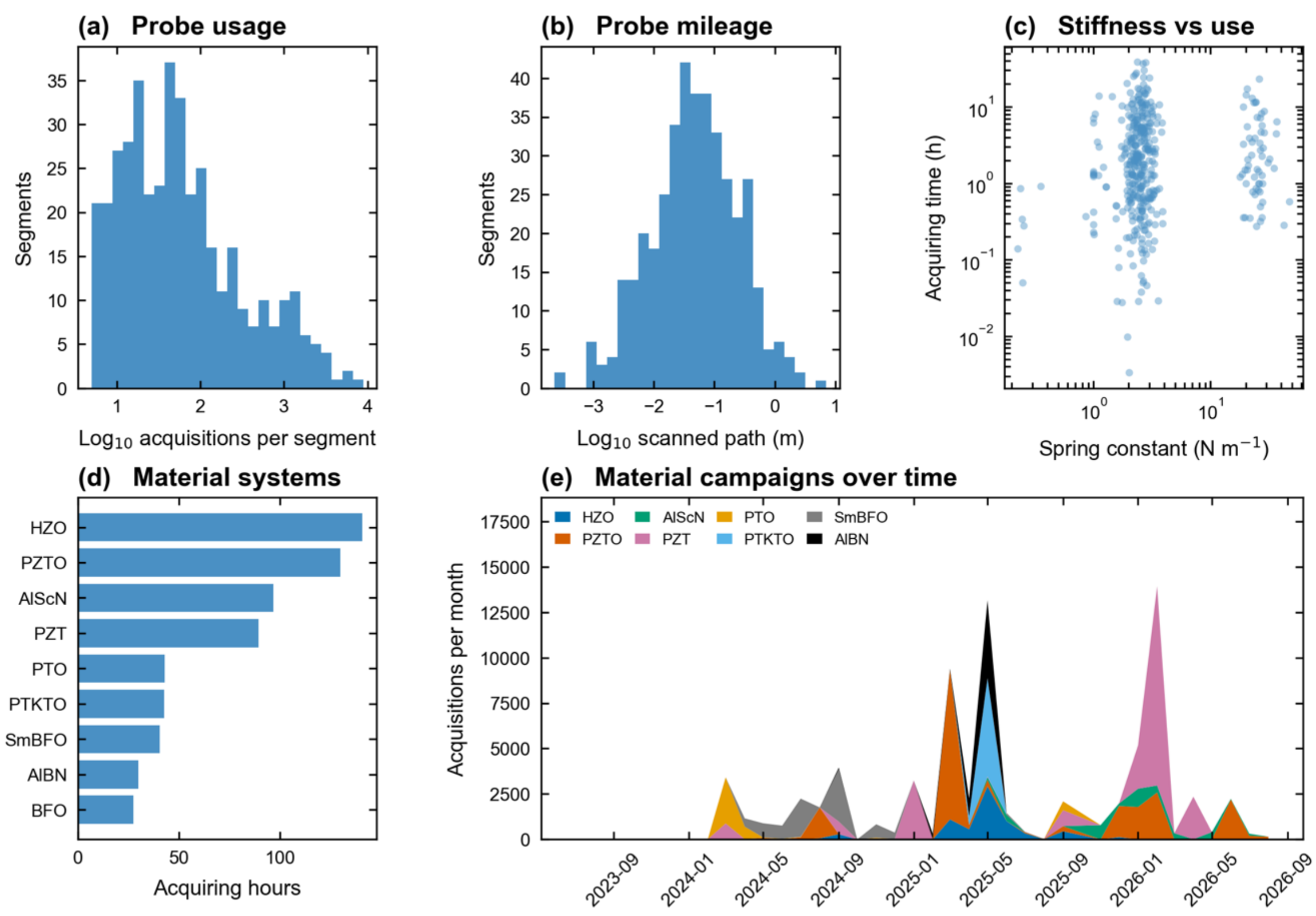

**Figure 3. Sample and probe state across the archive.** (a–b) Acquisitions and scanned path per inferred probe segment, both approximately log-normal over three decades. (c) Segment stiffness against accumulated use. (d) Acquiring hours by material system, from folder naming. (e) Material campaigns over time, showing the block structure of laboratory work.

### 3C. The instrument trace: latent probe and calibration states

Instrument metadata contains latent physical structure that is not explicit in a file hierarchy. This is particularly important in scanning probe microscopy, where the cantilever and tip are part of the measurement system. The archive does not contain a persistent probe identifier or a definitive 'tip changed' event, but it does preserve spring constant, inverse optical-lever sensitivity (InvOLS), amplitude InvOLS, resonance/drive frequency, phase offsets, and related calibration variables. We therefore represent a partial probe/calibration state as

$$P_t = [k_t,\ InvOLS_t,\ AmpInvOLS_t,\ f_drive_{,t},\ \ldots] \quad (1)$$

Contiguous fingerprinting of spring constant and optical-lever calibration identified 360 calibration epochs, and therefore 359 candidate transitions, at a 25 % relative-change criterion (Figure 4a); relaxing it to 15 % yields 489 segments. These are candidate probe/calibration changes rather than confirmed cantilever replacements. Recalibration, contact-state changes, probe replacement, or metadata anomalies can produce similar discontinuities, and a probe replaced by a nominally identical one is invisible, so the count is a lower bound.

Segment statistics nonetheless behave physically. The median inferred probe collects 45 acquisitions over 1.9 acquiring hours and 0.05 m of scanned path, and 62.4 % of segments span more than one session. (Figure 3a–c), so wear accumulates across days and the probe timeline must be built corpus-wide rather than per session. Both usage distributions are heavy-tailed and approximately log-normal (Figure 3a,b): across the 410 segments carrying at least five acquisitions the interquartile range spans 16-146 acquisitions and 0.015–0.145 m of scanned path, while the extremes reach 8,850 acquisitions and 7.1 m, and the busiest tenth of segments accounts for 70 % of all acquisitions and 61 % of the total scanned path. Cantilever stiffness is bimodal, with 85 % of segments near 2.5 N $m^{-1}$ and 15 % near 25 N $m^{-1}$, but accumulated use is statistically independent of it (Spearman $\rho = 0.04$ between log stiffness and log acquisitions, $p = 0.44$; median 48 and 38 acquisitions for the soft and stiff populations respectively). Based on this data, how long a probe remains in service is therefore set by what the campaign required rather than by the cantilever itself (Figure 3c), which is consistent with these being calibration epochs rather than confirmed probe lifetimes – a probe kept through a short campaign and one retired after a long one are indistinguishable in this representation.

The long archive also reveals several dense drive/contact-resonance regimes near 298, 319, 345, 361, 379, 409, and 441 kHz together with a heterogeneous high-frequency regime near 663 kHz (Figure 4b). Based on the experience, the high-frequency regime is attributed to the first lateral (cantilever twisting) resonance, that was used for the lateral DART PFM. This structure is important for interpreting environmental correlations. When frequency is plotted against recorded temperature without conditioning, the bands can be mistaken for a broad thermal relation. Within the dense lower-frequency regimes, apparent linear slopes range only from approximately −0.9 to +0.8 kHz per degree C and all have $R^2 \leq 0.06$. Thus latent probe/contact state dominates the observed frequency variation, and environmental effects should be estimated only after state separation. With the campaign confound of Section 3B, this is the second correction that must be applied before operator behavior is read from the trace.

Spatial coordinates provide another latent operational state (Figure 4 c,d). Recorded macroscopic stage moves range from stationary sessions to days containing hundreds of position changes, while local scan-center motion captures fine spatial exploration. These trajectories distinguish between stationary repeated spectroscopy, sample navigation, coarse-to-fine exploration, and global-local closed loops even before raw images are interpreted. Because stage coordinates can also jump during sample or chuck changes, the trace should eventually identify the reference frame and sample state explicitly rather than interpreting coordinate differences blindly as physical travel.

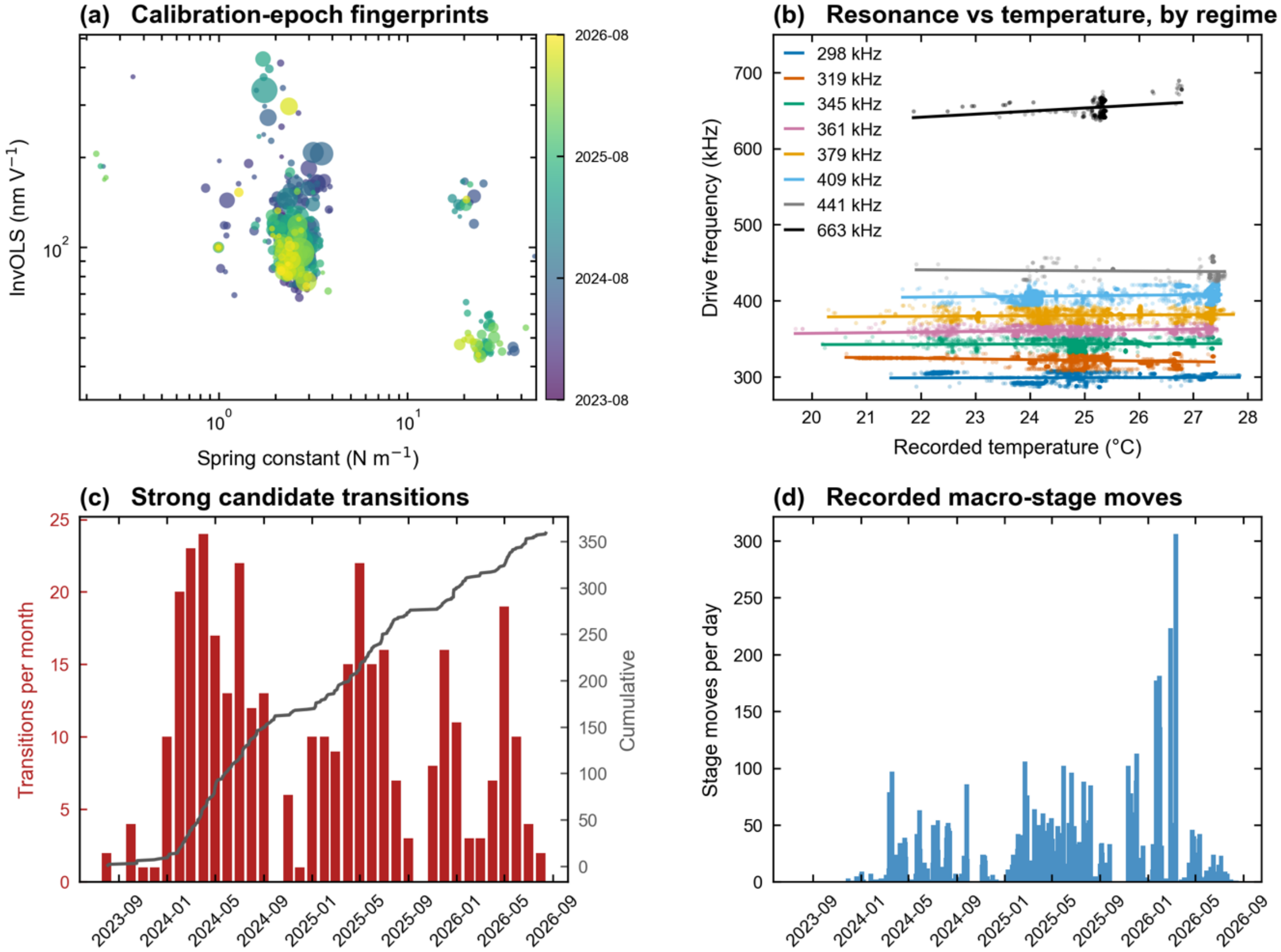

**Figure 4. Latent instrument, probe, and spatial states.** (a) Calibration-epoch fingerprints in InvOLS-spring-constant space; extreme calibration outliers are omitted from the displayed range but retained in the trace. (b) Drive/contact-resonance frequency versus recorded temperature separated into major operating regimes, showing that regime identity dominates the apparent temperature relation. (c) Timeline of strong candidate probe/calibration transitions. (d) Recorded macro-stage moves per day, illustrating the evolving spatial component of instrument use.

**3D. The decision trace at session scale: experimental complexity and operating strategy**

With sample and instrument state established, the remaining structure in the trace is behavioral. An Instrument Trace should distinguish a simple repeated measurement from an experiment that changes modalities, positions, waveforms, controls, or hierarchical programs. We therefore treat experimental complexity as multidimensional rather than as a proxy for scientific value. For each substantive session we calculate modality, spatial, waveform, control-path, sequence, and hierarchy descriptors. Each component is percentile-normalized over the archive and their mean is used only as a convenient composite visualization.

$$C = [C(modality), C(spatial), C(waveform), C(control), C(sequence), C(hierarchy)] \quad (2)$$

Principal-component analysis of the six standardized descriptors gives a complementary, non-arbitrary view of the session space (Figure 5a). PC1 explains 31.9 % of the variance and loads most strongly on sequence (0.638), modality (0.594), and waveform (0.377) complexity; it therefore acts primarily as a programmatic/operational-complexity axis. PC2 explains 23.8 % and is dominated by spatial (0.691), hierarchical (0.506), and control-path (0.435) complexity, separating spatially structured and global-local workflows from purely waveform-rich programs. The distribution changes from month to month but shows no monotonic progression, supporting the interpretation that complexity is campaign-specific rather than a simple measure of instrument maturity — which is consistent with, and partly explained by, the campaign block structure of Section 3B.

Experimental complexity should also be separated from control aggressiveness. For a recorded control vector $u_t$, we define the normalized size of a step using robust empirical scales $s_k$ for each control variable.

$$A_t = [\Sigma_k ((u_{k,t+1} - u_{k,t}) / s_k)^2]^{1t2} \quad (3)$$

We separately define control activity as the number of recorded control changes per session hour, where a recorded control change is any difference in the operator control vector including drive amplitude, amplitude and deflection setpoints, integral gains, scan geometry, waveform parameters and tip bias between two consecutive events in the same session. Drive and DART frequencies are excluded, because resonance tracking moves them without operator action. On this definition 610 of the 833 substantive sessions contain at least three recorded control changes. Median normalized step magnitude is 0.62 robust units and median recorded activity is 5.32 changes $h^{-1}$ (Figure 5d). Complexity and aggressiveness are effectively uncorrelated (Spearman $\rho = -0.078$, $p = 0.053$) (Figure 5b,c). This supports the distinction between these descriptors: a complex experiment need not be aggressive, a careful operator can still work rapidly, and very slow parameter changes are not automatically evidence of good performance. A future operator-carefulness metric must be conditioned on sample, probe, experiment type, and outcome quality and will require logging unsaved tuning operations — a requirement that Section 3F makes quantitative.

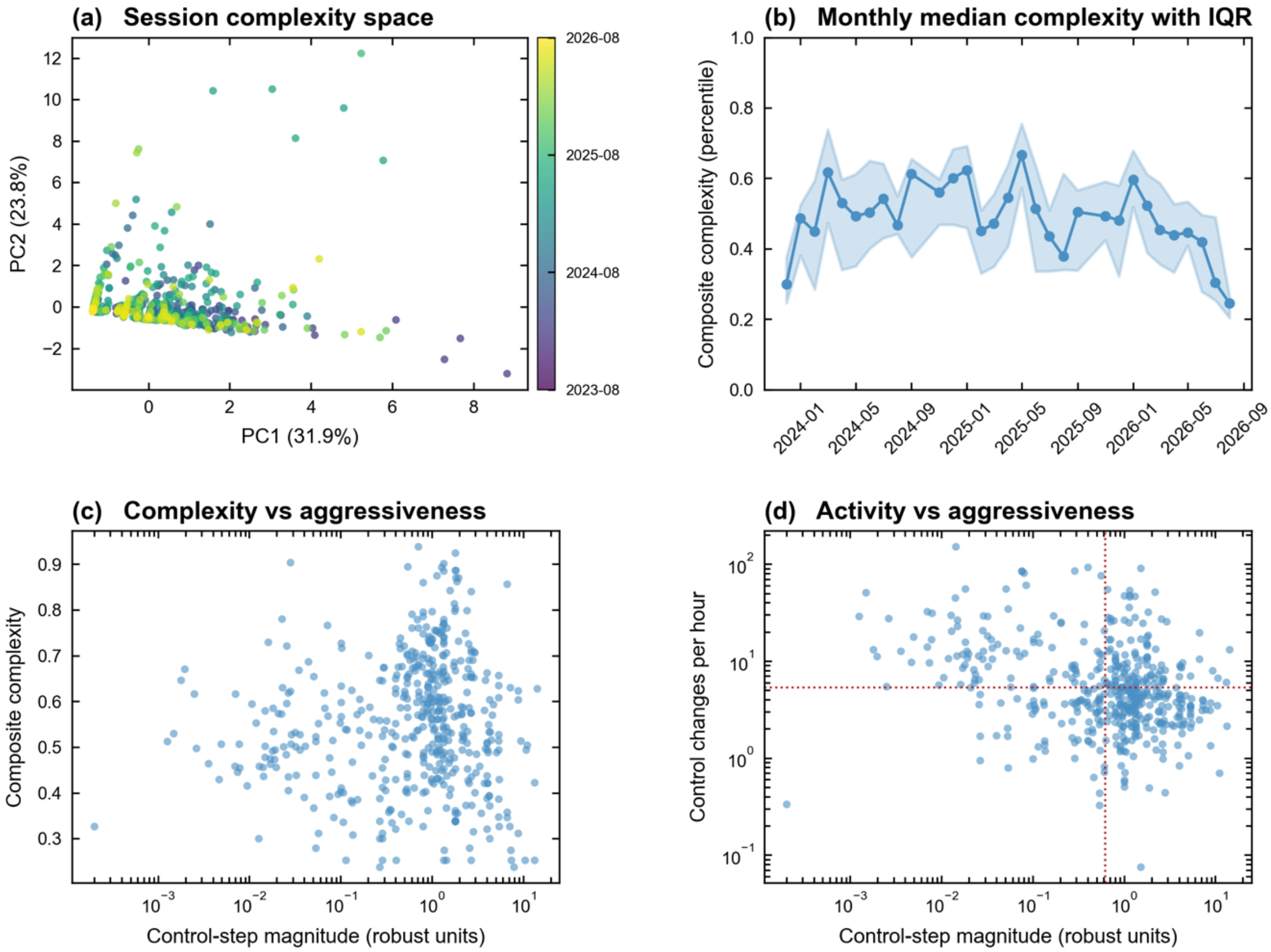

**Figure 5. Experimental complexity and operating strategy.** (a) PCA representation of 833 substantive sessions using six standardized complexity descriptors; color denotes time. PC1 explains 31.9% and PC2 23.8% of the descriptor variance. (b) Monthly median composite complexity with the interquartile range, showing campaign-dependent rather than monotonic evolution. (c) Composite complexity versus recorded control-step aggressiveness. (d) Recorded control-change activity versus aggressiveness; the logarithmic vertical scale exposes the main distribution while retaining rare large steps.

### 3E. The decision trace at event scale: experimental grammar

Chronological ordering contains information beyond individual metadata values. We therefore map the event stream to a small symbolic vocabulary: G denotes an explicitly identified global image, I an image, S simple spectroscopy, and W programmed or complex spectroscopy. The direct-follow matrix is highly structured. After an image, another image follows with probability 0.73, simple spectroscopy with 0.14, and programmed spectroscopy with 0.13. Simple spectroscopy follows itself with probability 0.59, while programmed spectroscopy is especially persistent, with $P(W_{t+1} \mid W_t) = 0.86$ (Figure 6a). The trace therefore captures a grammar of experimental behavior rather than an unordered collection of files.

A finer vocabulary makes this grammar interpretable as decisions rather than event types (Figure 6b,c). Each acquisition is labelled by the change relative to its predecessor. Five states describe imaging: survey, the first frame of a session; zoom out and zoom in, a scan size respectively above 1.25 and below 0.80 of the previous frame; repeat, the same frame re-acquired unchanged; and retry, the same frame re-acquired with control parameters changed. Relocate marks a frame center moved by more than half a frame width, and two states describe spectroscopy, point and grid, according to whether the acquisition sits inside a machine-paced batch. The fitted first-order transition model reproduces the marginal occupancy of every state to within 0.013 (Figure 6b,c). The corresponding decision graph is shown in Figure 6d.It is computed over the 39,704 acquisitions in the 604 presumably manually paced PFM sessions, the remainder of the archive being assigned to machine-paced operation by the timing criterion described in Methods.

The chain separates into two sub-loops of very unequal cost. The five imaging states carry only 27.9 % of acquisitions but 79.8 % of acquiring time, because an image takes minutes where a spectroscopic loop takes seconds. Here, median 256 s for a survey and 128 s for a zoom against 6 s for point spectroscopy. The two spectroscopy states are the mirror image, 63.2 % of acquisitions and 13.0 % of time, and they are strongly absorbing, with self-transition probabilities of 0.68 and 0.80; the remaining 8.9 % of acquisitions and 7.2 % of acquiring time correspond to relocation. Counting files and counting instrument time therefore give almost opposite pictures of the same chain.

Within the imaging loop, the operator strategy is bracketing rather than descending. Zoom out is followed by zoom in with probability 0.67, and zoom in returns to zoom out with 0.19, so magnification oscillates around a feature instead of progressing monotonically from coarse to fine. Repeat is the most persistent imaging state, self-transitioning at 0.51, which is deliberate re-imaging of an unchanged frame for drift tracking or waiting for the tip to settle, rather than an accident. Once inside the imaging loop the trajectory tends to stay there: the probability of remaining among imaging states is 0.92 from zoom out, 0.77 from repeat and 0.74 from survey.

**Relocate** is the hinge between the two loops, and it is almost entirely one-directional. After moving the frame, the next acquisition is spectroscopy with probability 0.90 and an image with probability 0.05. Moving is almost never an end in itself; it is the operator positioning for a measurement, and the imaging loop is what precedes rather than follows it.

The expensive state is the **retry**. It takes 18.8 % of acquisitions but 31.8 % of acquiring time - more than any other decision, and more than half again the next largest, repeat at 20.5 %. Retry is also the only imaging state that readily leaves the imaging loop, staying within it with probability just 0.25: it exits to spectroscopy at 0.29 and, tellingly, to relocate at 0.33, the signature of an operator who tries a few parameter sets, fails to fix the image, and abandons the spot. Retry is reached from both loops - from repeat at 0.18, from point spectroscopy at 0.27 and from grid spectroscopy at 0.19 - so it functions as the chain's general-purpose corrective state. Every retry encodes a quality judgement that the archive registers and does not explain.

The archive also contains 7309 repeated same-program, same-position spectroscopy chains. The median chain contains two repetitions, but the longest contains 4096. Only 10.5 % of repeated chains change spectroscopy amplitude and only 1.8 % change frequency, demonstrating that much of the spectroscopy history represents fixed-policy execution rather than visible parameter optimization. This distinction matters when comparing autonomy levels: repeated automated execution can be operationally efficient while containing little new decision agency.

The grammar nonetheless remains a description of what happened next, not of why an action was selected. It is recovered from saved files alone with no prior specification of the decision vocabulary, which is what makes it a candidate basis for the common action semantics of Section 4.3 – but the judgement between an observation and the action that follows it is absent at every point in the chain.

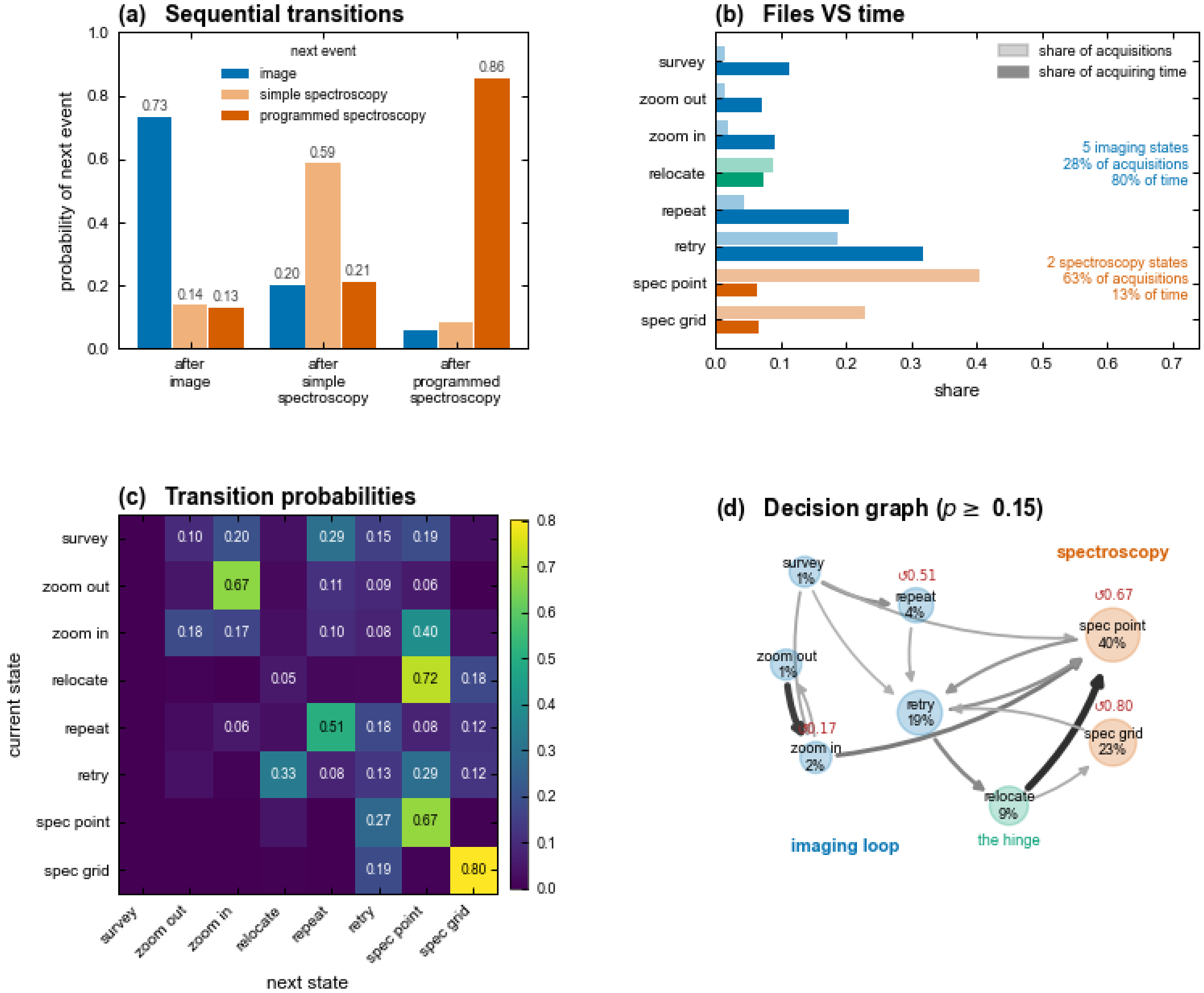

**Figure 6. Decision grammar of the manual PFM experiment.** (a) Probability that an image, simple spectroscopy, or programmed spectroscopy is followed by each event type. (b) Share of acquisitions (light) against share of acquiring time (solid) per decision state. (c) Row-stochastic transition matrix, values below 0.05 suppressed. (d) The same chain as a graph ($p \geq 0.15$), laid out as imaging (blue), spectroscopy (orange), and relocation (green) as the hinge between them; node area is proportional to share of acquisitions, self-transitions in red.

### 3F. The decision trace at control scale: tuning sequences

The retry state of Section 3E can be resolved one level further, to the individual recorded control change. We detect a candidate tuning operation as a transition between two successive saved images in the same session and approximately the same field of view, with a significant change in at least one of eight controls: amplitude setpoint (A), AC drive (V), DART frequency width (W), DART integral gain (D), deflection setpoint (S), imaging integral gain (I), scan rate (R), and DC tip bias (B). Drive frequency is excluded because DART tracking can change it without operator action. A tuning episode is defined as a contiguous run of such operations. We use a 5% relative-change threshold together with a small control-specific absolute floor; sensitivity tests at 2%, 5%, and 10% produce essentially the same statistics.

Recorded tuning is typically shallow but strongly heavy-tailed (Figure 7a). The median episode contains a single operation across material systems, acquisition modes, and manual/autonomous classes, whereas the upper tail varies substantially between strata (Figure 7b).

The most pronounced example is AlScN, for which the 95th-percentile episode approaches 50 operations, while PZT and PZTO remain much shallower. Manual and autonomous subsets show similar typical behavior, with a median of one operation in both cases. Thus the major heterogeneity lies not in the typical episode but in the relatively small population of long sequences.

Episode length alone does not tell us how much tuning effort was required. A tuning episode is a sequence of one or more consecutive control-change operations, and different episodes can represent very different kinds of instrument behavior. We therefore separate single-step episodes from multi-step sequences and classify the latter according to their internal pattern as local tuning, periodic switching, monotonic sweeps, or calibration blocks (Figure 7c). Most detected episodes consist of only a single operation, while most multi-step episodes correspond to local tuning. Periodic switching occurs in relatively few episodes, but these episodes contain many repeated control changes and therefore account for nearly half of all detected operations. Thus, a long episode does not necessarily mean that the operator spent a long time searching for suitable imaging conditions; it may instead reflect a regular, structured sequence of instrument states. These different sequence types must therefore be separated before episode length or the number of control changes can be used as a measure of tuning effort.

At the level of the controls themselves, changes recur in stable composite bundles rather than as independent scalar adjustments. Transitions among the most frequent bundles show pronounced diagonal structure (Figure 7d): once a control subspace such as AI, AVWDS, or VRB is engaged, subsequent adjustments frequently remain within the same bundle. The trace therefore contains a data-derived vocabulary of composite experimental actions above the level of individual instrument parameters, without requiring those actions to be specified in advance. This provides an empirical basis for the common action semantics introduced in Section 4.3.

Tuning also forms part of a recurrent experimental cycle rather than an isolated correction. We identify 1,142 tuning → stable acquisition → tuning round trips with a median duration of 17.3 min (Figure 7e). Across these cycles, approximately 20% of the wall-clock time is spent in tuning, 69% in stable acquisition, and 11% in navigation. The operational pattern is therefore one of intermittent intervention embedded within substantially longer periods of stable measurement: parameters are adjusted, evidence is acquired under the resulting state, and tuning is re-entered when further intervention becomes necessary.

Finally, we asked whether recorded tuning activity changes before an abort (Figure 7f). For 131 abort events, we calculated the number of detected control-change operations relative to the number of acquisitions at different times before the abort command. Overall, tuning activity is higher several hours before the event and becomes lower as the abort approaches, although the trend is not monotonic. This decrease should not be interpreted directly as a warning signal for an approaching failure. Microscope tuning is typically concentrated near the beginning of an experiment, followed by longer periods of more stable acquisition with fewer parameter changes. If aborts tend to occur later in a session, a similar decrease would therefore appear even without a direct connection to the abort itself. The present result shows that recorded tuning activity changes systematically over the period preceding aborts, but determining whether this behavior is specific to failure will require comparison with normal, non-abort periods at similar stages of an experimental session.

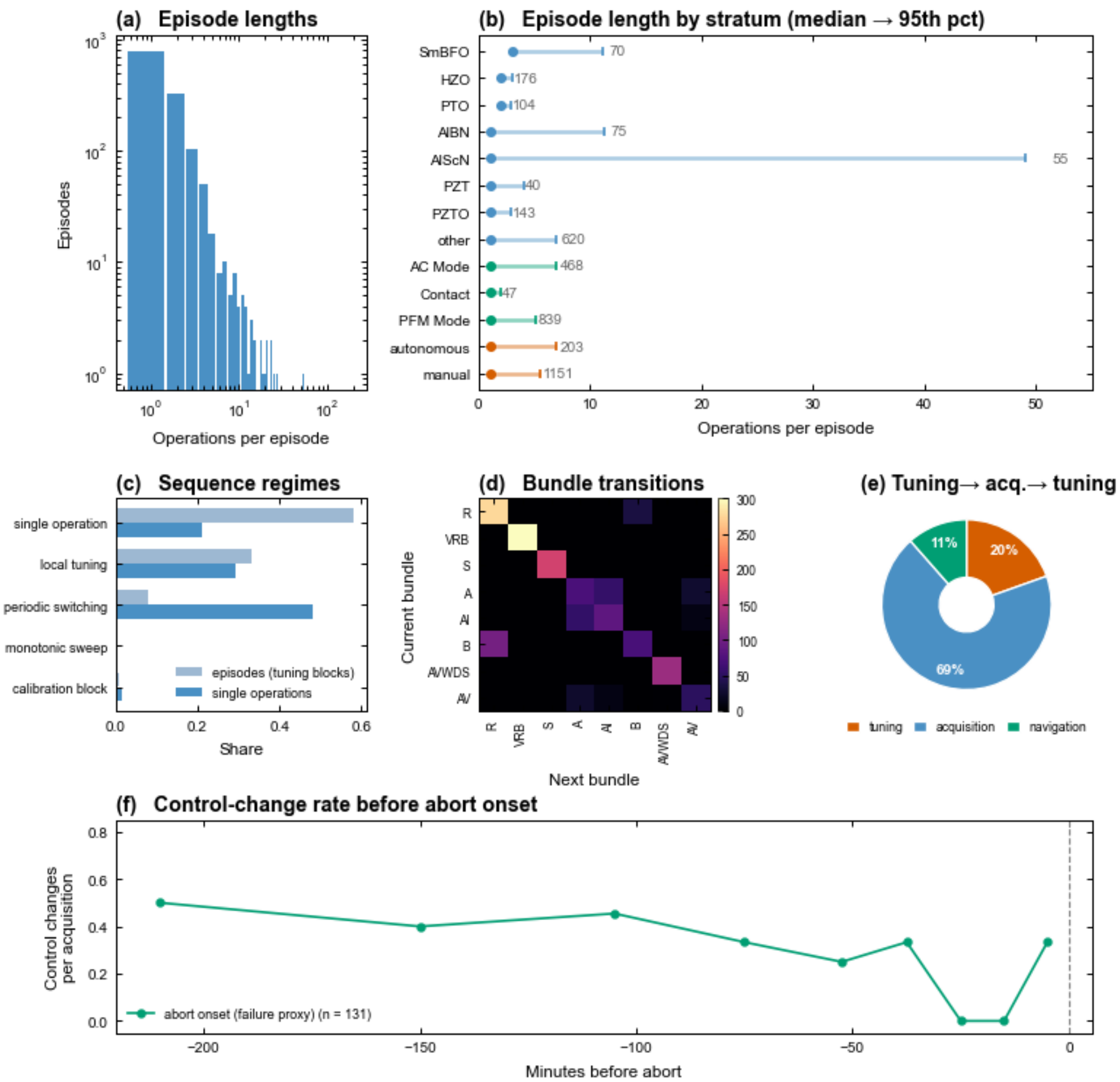

**Figure 7. Recorded tuning structure and pre-abort control activity.** (a) Distribution of tuning-episode lengths. (b) Median and 95th-percentile episode length across material systems, acquisition modes, and manual/autonomous operation. (c) Share of tuning episodes and individual control operations assigned to different sequence regimes, separating single-step corrections, local tuning, periodic switching, monotonic sweeps, and calibration blocks. (d) Transitions between recurrent multi-parameter control bundles. (e) Wall-clock composition of tuning → acquisition → tuning round trips, showing that most time is spent in stable acquisition rather than tuning or navigation. (f) Recorded control-change rate before abort onset, expressed as detected tuning operations per acquisition.

## 4. From Instrument Traces to Decision Traces

The retrospective results show that existing scientific archives already contain substantial information about how experiments were performed. They also show what cannot be recovered reliably after the fact, and the boundary is the same at every scale examined in Section 3: the archive records the action and its outcome, and not the judgement between them. Future active instruments should therefore treat trace generation as part of normal scientific operation rather than an optional diagnostic. Three questions follow: what a single instrument must write down, what it must write down about the choices made, and what must be agreed across instruments for any of it to transfer.

### 4.1 What an active instrument must record

A prospective trace requires a universal core plus instrument-specific extensions. The universal core should persist project, campaign, sample, sample state, session, instrument, probe where relevant, operator or controller, event time, action, outcome, and error state. Human operator identities should remain available internally for longitudinal training and facility support but be pseudonymized for shared datasets and reports. Every action should identify its control source, distinguishing human control, native feedback, fixed scripts, optimizers, learned policies, agents, and mixed human-AI operation. The physical extension is necessarily modality specific: PFM requires probe, resonance, calibration, excitation and switching-program variables; KPFM, STEM, XRD, synthesis robots, and other systems require different state descriptors. Sample and campaign identity deserve the same treatment, since Section 3B shows that material and epoch are otherwise confounded beyond retrospective repair.

Exogenous state should also be first-class data. Temperature, humidity, vibration, acoustic noise, power/HVAC state, and exceptional external events can influence outcome quality while remaining invisible to the controller. The Nega et al. humidity result [44] illustrates the scientific cost of treating such variables as unrecorded noise; Section 3C illustrates the converse failure, in which an apparent temperature dependence dissolves once latent probe state is conditioned out. Importantly, all actions should be logged, not only successful files. Unsaved tuning, preview scans, aborted acquisitions, retries, failed calibrations, warnings, interlocks, and recovery events are precisely the records needed to learn careful operation and impending failure.

The present archive records acquisition time much better than total instrument-use time. A future time trace should distinguish sample/setup, tuning, navigation, acquisition, human reasoning, agent reasoning, data transfer, waiting, and recovery. In a cloud-controlled workflow the microscope may wait for the agent, the agent may wait for the microscope, or both may wait for data transfer or external analysis. These latencies cannot generally be parallelized away because later decisions depend on newly acquired evidence. Login/logout and instrument-application state should therefore be integrated with file histories rather than assuming that the absence of a saved acquisition represents idle time.

The trace should also record whether the measurement was any good, which the current metadata does not. Raw image and spectroscopy arrays can add signal-to-noise, contrast, dynamic range, clipping, line artifacts, drift, creep, feedback instability, forward/retrace consistency, feature density, spectral fit residuals, loop quality, and domain-specific descriptors; prior microscopy data-analysis work illustrates how physically constrained, statistical, and representation-learning descriptors can be extracted reproducibly from such payloads [48-52]. Artifact probability should be represented independently from visual quality. In SPM, image-based estimates of double-tip behavior or probe degradation can be combined with calibration history to define a functional probe state. This enables conditional analyses such as (experiment complexity, control strategy, probe state) $\rightarrow$ measurement quality, and eventually identifies which materials or control policies accelerate probe degradation. Adding this data also converts the retry of Section 3E into a supervised dataset, since the rejected frame, the correction applied, and the eventual accept-or-abandon outcome are all already recorded.

Critically, the outcome value should be allowed to evolve after acquisition. A measurement may later become a publication figure, support a reusable dataset, train an operator, enable a subsequent project, or acquire value through reanalysis with better algorithms. The original data and trace remain fixed, while derived quality metrics, scientific interpretation, and downstream value can be updated. This distinction is important for facility assessment: data persistence is immediate, but scientific value is often delayed.

Recording control source per event also settles how autonomy should be described. The same physical microscope can host conventional human imaging, fixed-policy scripts, Bayesian optimization, learned policies, human-agent collaboration, and fully agent-controlled campaigns, and the archive supports this view: closed-loop motifs occur strongly in particular campaigns but do not define the instrument as a whole. Autonomy should therefore be treated as a time- and campaign-dependent property rather than a binary instrument label, and, where the trace permits, as a measured rather than an asserted one. This is consistent with broader SDL performance frameworks, which distinguish degree of autonomy from throughput, precision, accessible parameter space, and optimization performance [53].

**4.2 Decision traces and scientific agency**

Decision provenance becomes essential as control shifts from fixed automation toward optimizers and agents. For AI-controlled experiments, the trace can preserve not only the chosen command but the goal, evidence, reward or objective, candidate actions, uncertainty or utility, selected action, latency, safety filtering, outcome, and human override. A compact decision object can therefore be written as

$$D_t = \{G_t, E_t, R_t, \mathcal{A}_t, U_t, a_t, \tau_t, override, outcome\} \quad (4)$$

Human decision traces should be lighter weight. It is neither practical nor desirable to require narration of every knob movement, and Section 3 indicates where the boundary falls: the event-level grammar is compact and most tuning episodes are a single operation, so routine execution can be recorded without requiring narration of every step. Consequential choices, by contrast, can be recorded at campaign or project level: why a new spectroscopy program was introduced, why a region was selected, why the sample was modified, or why the experiment was stopped. This allows execution to be separated from agency. A fixed script can execute thousands of measurements with no new high-level decision, while one decision can redirect an entire project.

We propose GERD - Goal-Evidence-Reward-Decision - as a unit of scientific agency, defined so as to count the second quantity rather than the first. The PZTO1 session illustrates the arithmetic: 671 recorded events reduce to 215 bracketed observation-intervention-observation cycles under a single unchanged policy. An expert-operator week provides a practical human baseline, but GERD is not simply 40 h of instrument or labor time. A quantitative GERD comparison should be calibrated on matched human and agent tasks and should weight meaningful decisions by outcome quality, safety, and auditability.

**4.3 What must be standardized across instruments**

Portable autonomy requires a separation between the semantics of an experimental action and the proprietary implementation of a particular instrument. Across microscopes, operators repeatedly invoke a small set of higher-level operations - tune, navigate, image, zoom, acquire spectroscopy, manipulate, verify - even though vendors implement them differently. We therefore propose common action semantics, or a domain-specific hyperlanguage, above vendor APIs, consistent with earlier microscopy workflow and cloud/edge-control architectures [24, 25, 47]. The trace ontology records what happened; the hyperlanguage defines what a human or agent is allowed to ask the instrument to do. This preserves an open workflow layer without requiring manufacturers to expose proprietary low-level controller code. Such a vocabulary need not be imposed by fiat: the decision states of Section 3E and the parameter bundles of Section 3F were both derived from saved files with no prior specification of what the composite actions should be, and a hyperlanguage designed against empirically recurrent actions is more likely to be learnable and portable than one designed against the scalar control surface a vendor happens to expose.

Physical automation also needs an independent safety layer. A manufacturer-supported 'Guardian Angel' module can evaluate a proposed action against instrument state, probe state, sample class, safe operating envelopes, and accumulated failure history before execution. Rejected actions should be logged alongside executed ones. Such a layer is analogous to a safety interlock but can become state- and context-aware as federated operational data accumulate. Autonomous STM tip conditioning already demonstrates that measurement data can identify a degraded probe state and trigger corrective action [31]; the same logic can be generalized to broader dangerous-state prediction.

Agreement on semantics is also what allows experience to move between facilities and is a prerequisite for cloud-connected or federated instrument operation [25, 47]. For a mature combination of material, instrument, probe, and workflow, the best prior information may already exist in the local trace. The cold-start problem appears when a new material, new probe family, or unfamiliar experimental goal is introduced. Federated traces can then provide experience from related instruments and facilities, allowing initial operating conditions and recommended carefulness to be conditioned on material descriptors, probe descriptors, instrument state, and goal. Federation does not require every tool to record identical physical parameters; it requires a shared core ontology plus domain-specific state/action extensions and common action semantics. Shared combinatorial libraries and standard benchmark workflows can serve as particularly useful transfer objects because many facilities explore overlapping chemical or structural spaces.

### 4.4 What complete traces make possible

A complete trace enables retrospective counterfactual analysis. Given a stored state, observation, action, and subsequent outcome, a digital twin or learned surrogate can ask what might have happened under alternative feasible actions one step ahead. Counterfactual replay can estimate decision regret, identify redundant high-cost measurements, and compare human, script, optimizer, and agent policies without repeating every physical experiment. The trace therefore becomes training data for the instrument's own digital twin, and the retry chains of Section 3E are the natural first target; this extends earlier proposals for edge/cloud microscopy workflows from data transport toward persistent operational memory [25, 47].

Benchmarking should accompany this capability. Standard reference or 'dummy' workflows of increasing complexity can test whether a controller can obtain a valid image, tune to a quality threshold, navigate to a target, perform observation-intervention-verification, explore an unknown sample, and recover from faults. Online virtual microscopes or a Microscopy Gym can hide the underlying simulator from the agent while exposing the same action semantics as the physical instrument. Published active-experiment studies should consequently provide not only data and code, but where possible the Instrument Trace, controller or policy, an executable reference workflow, and representative or synthetic data sufficient to exercise the workflow without the original hardware.

$$\textit{maximize over } \pi\textit{: scientific progress /(time or agency), subject to } P(\textit{damage or failure}|\pi) < \varepsilon \quad (5)$$

Taken together, these capabilities change what an instrument can be asked to optimize. A digital twin or federated experience model can recommend not only an initial parameter state but how rapidly controls should be changed. The objective is to maximize scientific progress per time or unit agency while keeping predicted damage or failure below an acceptable threshold.

## 5. Conclusions

Active and autonomous instruments change what constitutes a complete scientific record. FAIR data stewardship, instrumental metadata, and reproducible analysis workflows remain necessary, but once observations determine subsequent measurements an additional object becomes essential: the experimental trajectory. We introduce the Instrument Trace as a longitudinal representation of this trajectory and argue that it should ultimately be synchronized with Sample and Decision Traces.

The multiyear AFM/PFM archive demonstrates that substantial structure can already be recovered retrospectively from conventional files, at every scale of the hierarchy. We reconstruct overall instrument activity and schedule modes; the block structure of material campaigns in the sample stream; candidate probe/calibration states and environment-conditioned resonance regimes; session-level experimental complexity and control strategy; spatial navigation, spectroscopy programs and fixed-policy repetition; a structured experimental grammar of decisions dominated by the retry; and, within the retry, a stable vocabulary of composite parameter actions. The PZTO example shows that an instrument archive can reveal a clean global-local observation-intervention hierarchy without access to controller source code. At the same time, the analysis exposes the limits of retrospective reconstruction, and they are the same limits at every scale: saved files incompletely record tuning and failed operations, cannot uniquely identify probe changes, and generally contain no representation of decision rationale. Richly recoverable action structure sits alongside entirely absent judgement.

Future active instruments should therefore preserve hierarchy, identities, complete timing, successful and failed actions, probe and sample state, exogenous variables, errors and recovery, controller source, and decision provenance. Raw measurements can then support quality, artifact, and tip-state analysis; retry chains can supply the supervised quality labels an autonomous agent currently lacks; longitudinal operator histories can support training rather than ranking; and safety modules can learn precursors to dangerous states. Common action semantics can make workflows portable above proprietary instrument implementations, while complete traces allow counterfactual replay, virtual-instrument benchmarking, and federated transfer of experience between facilities.

The resulting archive is not static in value. The original measurements and trace remain fixed, but analysis can evolve as better models emerge, operator training outcomes become known, data are reused, and scientific significance becomes clearer. An Instrument Trace therefore changes the archive from a repository of images and spectra into a longitudinal record of how scientific knowledge was acquired - and, increasingly, how humans and AI decided what to do next.

**Acknowledgements.** This work (SVK, BS, RL, and CC) is supported by the NSF PCL ATHENA, Award 2607469.

**6. Methods**

All analyses were based on metadata stored in the raw microscope files together with their organization in the workstation memory. Individual records were grouped into experimental sessions using a 60-min inactivity gap. To exclude very short or incomplete runs, only sessions containing at least three events and lasting at least 30 min were treated as substantive sessions. Because the microscope metadata do not directly record whether a session was controlled manually or automatically, we inferred the control mode from the timing of consecutive saved measurements. Here, an acquisition means one saved measurement event, such as an image or spectroscopy measurement. This timing provides a useful criterion because automated workflows usually repeat measurements with a relatively regular cadence, whereas human-operated experiments contain more variable delays associated with image inspection, parameter adjustment, navigation, and decision making. Intervals between consecutive acquisitions were grouped into 10%-wide logarithmic bins, and cadence repeatability was evaluated within a centered window of 11 acquisitions. An acquisition was considered machine-paced when at least 60% of the intervals fell within the two most populated timing bins. Two bins were used because automated workflows can alternate between two characteristic acquisition times while still remaining highly regular.

Contiguous machine-paced sequences containing at least 25 acquisitions were identified as automated batches. A session containing at least 10 acquisitions was classified as automated when at least 50% of its acquisitions belonged to such batches and the operator-controlled parameters changed in no more than 40% of consecutive steps. This second condition helps distinguish repeated automated acquisition from human tuning, where control parameters are changed more frequently. Sessions that did not satisfy these criteria were classified as manual. The resulting distinction should therefore be interpreted as a separation between machine-paced and predominantly human-paced operation, rather than as a direct identification of a particular automation algorithm or agent.

A tuning operation was defined as a transition between two successive saved images in the same session and approximately the same field of view, accompanied by a significant change in at least one of eight controls: amplitude setpoint, AC drive, DART frequency width, DART integral gain, deflection setpoint, imaging integral gain, scan rate, or DC tip bias. The scan-size ratio was required to remain within 1.25 and the frame center could not move by more than half a frame width, preventing relocation events from being counted as tuning. A 5% relative-change threshold with a small control-specific absolute floor was used, and the analysis was checked at 2% and 10%. Consecutive tuning operations were grouped into tuning episodes.

Multi-operation episodes were then classified according to their internal structure before being interpreted as tuning effort. Episodes were identified as calibration blocks when calibration constants changed substantially, as periodic switching when a small number of control bundles repeatedly recurred, as monotonic sweeps when a control changed systematically in one direction, and as local tuning otherwise. For the phase analysis, events were further assigned to navigation, tuning, or stable acquisition, allowing reconstruction of tuning → acquisition → tuning round trips and their wall-clock composition. Pre-abort control activity was evaluated from the number of detected tuning operations per acquisition in successive time windows before abort onset. Because the archive contains only saved events, these analyses describe recorded instrument behavior rather than the full sequence of human or algorithmic decisions.

## References

1. M. D. Wilkinson, Scientific Data **3**, 160018 (2016).
2. J. Jumper, R. Evans and A. Pritzel, Nature **596**, 583-589 (2021).
3. M. Baek, F. DiMaio and I. Anishchenko, Science **373**, 871-876 (2021).
4. A. Jain, S. P. Ong and G. Hautier, APL Materials **1**, 011002 (2013).
5. S. Kirklin, J. E. Saal and B. Meredig, npj Computational Materials **1**, 15010 (2015).
6. J. E. Saal, S. Kirklin, M. Aykol, B. Meredig and C. Wolverton, JOM **65**, 1501-1509 (2013).
7. S. Curtarolo, W. Setyawan and G. L. W. Hart, Computational Materials Science **58**, 218-226 (2012).
8. K. Choudhary, K. F. Garrity and A. C. E. Reid, npj Computational Materials **6**, 173 (2020).
9. C. Draxl and M. Scheffler, MRS Bulletin **43**, 676-682 (2018).
10. A. Talapatra, S. Boluki, P. Honarmandi and R. Arroyave, Frontiers in Materials **6**, 82 (2019).
11. L. Ward, A. Dunn and A. Faghaninia, Computational Materials Science **152**, 60-69 (2018).
12. M. Ziatdinov, C. T. Nelson and S. V. Kalinin, Microscopy and Microanalysis **25**, 170-171 (2019).
13. S. Somnath, A. Belianinov, S. V. Kalinin and S. Jesse, Advanced Structural and Chemical Imaging **4**, 3 (2018).
14. B. H. Savitzky, S. E. Zeltmann, L. A. Hughes and C. Ophus, Microscopy and Microanalysis **27**, 712-743 (2021).
15. M. Ziatdinov, A. Ghosh, C. Y. Wong and S. V. Kalinin, Nature Machine Intelligence **4**, 1101-1112 (2022).
16. C. Goble, S. Cohen-Boulakia and S. Soiland-Reyes, Data Intelligence **2**, 108-121 (2020).
17. A. S. Wagner, Scientific Data **9**, 80 (2022).
18. W. A. Shewhart, *Economic Control of Quality of Manufactured Product*. (D. Van Nostrand Company, New York, 1931).
19. S. J. Qin, Journal of Chemometrics **17**, 480-502 (2003).
20. V. Venkatasubramanian, R. Rengaswamy, S. N. Kavuri and K. Yin, Computers & Chemical Engineering **27**, 327-346 (2003).
21. A. K. S. Jardine, D. Lin and D. Banjevic, Mechanical Systems and Signal Processing **20**, 1483-1510 (2006).
22. W. M. P. van der Aalst, *Process Mining: Discovery, Conformance and Enhancement of Business Processes*. (Springer, Berlin Heidelberg, 2011).
23. S. V. Kalinin, M. Ziatdinov, J. Hinkle, S. Jesse, A. Ghosh, K. P. Kelley, A. R. Lupini, B. G. Sumpter and R. K. Vasudevan, ACS Nano **15**, 12604-12627 (2021).
24. S. V. Kalinin, M. Ziatdinov, M. Ahmadi, A. Ghosh, K. Roccapriore, Y. Liu and R. K. Vasudevan, Applied Physics Reviews **11**, 011314 (2024).
25. S. V. Kalinin, D. Mukherjee, K. Roccapriore, B. J. Blaiszik, A. Ghosh, M. A. Ziatdinov, A. Al-Najjar, C. Doty, S. Akers, N. S. Rao, J. C. Agar and S. R. Spurgeon, npj Computational Materials **9**, 227 (2023).
26. R. K. Vasudevan, K. P. Kelley, J. Hinkle, H. Funakubo, S. Jesse, S. V. Kalinin and M. Ziatdinov, ACS Nano **15**, 11253-11262 (2021).
27. Y. Liu, R. K. Vasudevan, K. P. Kelley, H. Funakubo, M. Ziatdinov and S. V. Kalinin, npj Computational Materials **9**, 34 (2023).
28. M. J. Olszta and S. R. Spurgeon, Microscopy and Microanalysis **28**, 1611-1621 (2022).
29. M. Ziatdinov, O. Dyck, S. Jesse and S. V. Kalinin, ACS Nano **11**, 12742-12752 (2017).

30. A. Krull, P. Hirsch, C. Rother, A. Schiffrin and C. Krull, Communications Physics **3**, 54 (2020).
31. M. Rashidi and R. A. Wolkow, ACS Nano **12**, 5185-5189 (2018).
32. E. A. Stach, B. L. DeCost and A. G. Kusne, Matter **4**, 2702-2726 (2021).
33. F. Hase, L. M. Roch and A. Aspuru-Guzik, Trends in Chemistry **1**, 282-291 (2019).
34. B. P. MacLeod, Science Advances **6**, eaaz8867 (2020).
35. B. Burger, Nature **583**, 237-241 (2020).
36. J. M. Granda, L. Donina, V. Dragone, D.-L. Long and L. Cronin, Nature **559**, 377-381 (2018).
37. B. J. Shields, Nature **590**, 89-96 (2021).
38. I. M. Pendleton, G. Cattabriga, Z. Li, M. A. Najeeb, S. A. Friedler, A. J. Norquist, E. M. Chan and J. Schrier, MRS Communications **9**, 846-859 (2019).
39. J. Bai, S. Mosbach, C. J. Taylor, D. Karan, K. F. Lee, S. D. Rihm, J. Akroyd, A. A. Lapkin and M. Kraft, Nature Communications **15**, 462 (2024).
40. A. G. Kusne, Nature Communications **11**, 5966 (2020).
41. N. J. Szymanski, B. Rendy and Y. Fei, Nature **624**, 86-91 (2023).
42. J. T. Rapp, B. J. Bremer and P. A. Romero, Nature Chemical Engineering **1**, 97-107 (2024).
43. A. A. Volk, R. W. Epps and D. T. Yonemoto, Nature Communications **14**, 1403 (2023).
44. P. W. Nega, Z. Li, V. Ghosh, J. Thapa, S. Sun, N. T. P. Hartono, M. A. N. Nellikkal, A. J. Norquist, T. Buonassisi, E. M. Chan and J. Schrier, Applied Physics Letters **119**, 041903 (2021).
45. M. Konnecke, F. A. Akeroyd and H. J. Bernstein, Journal of Applied Crystallography **48**, 301-305 (2015).
46. S.-A. Sansone, Nature Biotechnology **37**, 358-367 (2019).
47. S. V. Kalinin, K. Barakati, B. N. Slautin, A. C. Houston, U. Pratiush, G. Duscher, C. Cao, I. Sgouralis, S. R. Spurgeon, C. Yang and G. Channing, Microscopy Today **34**, 16-24 (2026).
48. S. V. Kalinin, B. G. Sumpter and R. K. Archibald, ACS Nano **10**, 9068-9086 (2016).
49. M. Ziatdinov, A. Maksov and S. V. Kalinin, npj Computational Materials **3**, 31 (2017).
50. R. Kannan, Advanced Structural and Chemical Imaging **4**, 6 (2018).
51. M. Ziatdinov, S. Jesse, R. K. Vasudevan and S. V. Kalinin, npj Computational Materials **6**, 21 (2020).
52. S. V. Kalinin, Science Advances **7**, eabd5084 (2021).
53. A. A. Volk and M. Abolhasani, Nature Communications **15**, 1378 (2024).